\documentclass[a4paper,11pt]{article}

\usepackage{pos}
\usepackage{lipsum} 
\usepackage{wrapfig}
\usepackage{subcaption}
\usepackage{enumitem}
\newlength{\bibitemsep}
\newlength{\bibparskip}
\let\oldthebibliography\thebibliography
\renewcommand\thebibliography[1]{%
  \oldthebibliography{#1}%
  \setlength{\parskip}{\bibitemsep}%
  \setlength{\itemsep}{\bibparskip}%
}

\title{Performance of the Prototype Station of the IceCube Surface Array Enhancement}

\ShortTitle{Performance of SAE station at IceCube}

\author{The IceCube Collaboration \\{\normalsize \normalfont(a complete list of authors can be found at the end of the proceedings)}\\}

\emailAdd{shefali@icecube.wisc.edu}

\abstract{
The prototype station of the Surface Array Enhancement at the IceCube Neutrino Observatory has been taking data in its final design since 2023. This station is part of the planned extension within the footprint of the existing surface array, IceTop. One station consists of 8 scintillator detectors, 3 radio antennas, and a central DAQ. 
The final upgrade of the scintillation detectors and their firmware at the prototype station has extended the dynamic range and increased the data-taking up-time, thereby expanding the observation window for air showers.
This contribution will discuss the performance of the upgraded prototype station after commissioning and its angular resolution capabilities when observing air showers with the scintillation detectors and in coincidence with IceTop. Furthermore, the integration of additional stations during the most recent deployment will be discussed.

\vspace{4mm}

{\bfseries Corresponding authors:}
Shefali$^{1*}$\\
{$^{1}$ \itshape Institute of Experimental Particle Physics, Karlsruhe Institute of Technology (KIT), Germany}\\
[4mm]
$^*$ Presenter
}

\ConferenceLogo{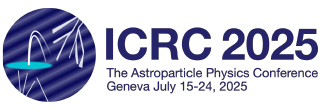}

\FullConference{39th International Cosmic Ray Conference (ICRC2025)\\
 15–24 July 2025\\
Geneva, Switzerland\\}

\begin{document}
\maketitle

\section{Introduction}
The IceCube Neutrino Observatory is a hybrid detector encompassing a volume of one cubic kilometer of Antarctic ice at the South Pole. It consists of a deep in-ice array of optical sensors and a surface array known as IceTop. Since its full deployment in 2011, IceTop has played a vital role in vetoing atmospheric backgrounds for high-energy neutrino detection and has proven to be an effective instrument for cosmic-ray studies, providing key measurements of the cosmic-ray flux in the Southern Hemisphere~\cite{Soldin:2022evu} in the PeV primary energy range. A major challenge for IceTop is non-uniform snow accumulation on surface detectors over time, which attenuates signals and increases the trigger threshold as well as uncertainties in energy and mass reconstruction.

\begin{wrapfigure}{r}{0.47\textwidth}
    \centering
    \vspace{-0.4cm}
    \includegraphics[width=\linewidth]{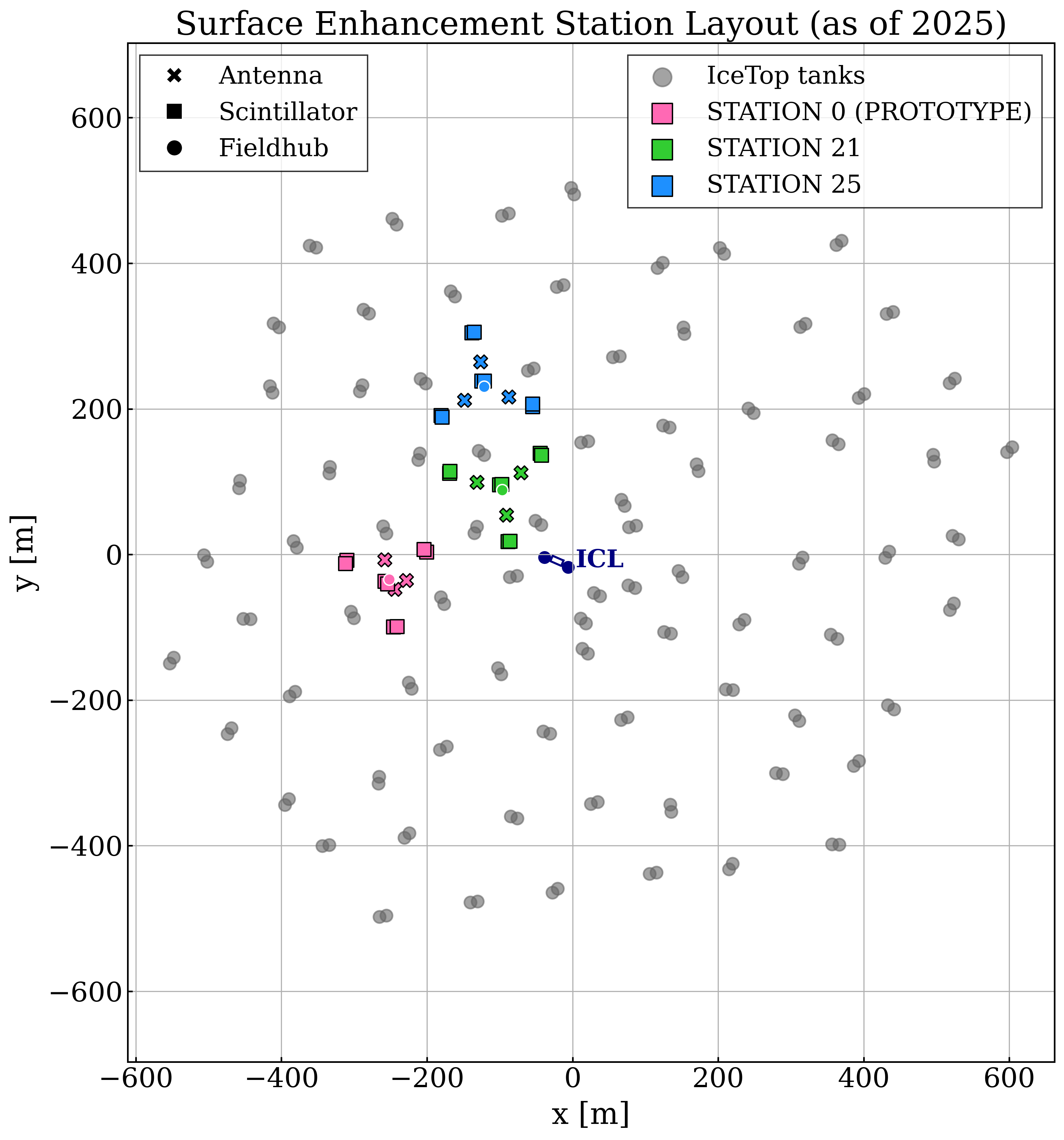}
    \vspace{-0.6cm}
    \caption{Layout of the deployed SAE stations at the South Pole. Grey circles indicate IceTop stations corresponding to in-ice strings.}
    \label{layout}
    \vspace{-0.5cm}
\end{wrapfigure}

While ongoing efforts aim to correct for this~\cite{Rawlins2023}, accurate treatment remains difficult due to sparse and infrequent snow depth measurements and particle-dependent attenuation, muons being less affected than electromagnetic particles. To address these limitations and to benefit from multiple cosmic-ray detection channels, the Surface Array Enhancement (SAE)~\cite{Haungs_2019} has been proposed. The SAE will consist of hybrid detector stations deployed along the footprint of IceTop. Each station includes eight scintillation detectors, three radio antennas, and a central data acquisition (DAQ) system. At present, three full SAE stations are deployed and operational at the South Pole. The prototype station, also referred to as Station 0~\cite{shefali2024statusplansinstrumentationicecube}, was first deployed in 2020 and upgraded to its final configuration in 2023. In the most recent deployment season, two additional stations, named Stations 21 and 25 (named according to the planned full-array geometry), were successfully installed. Figure~\ref{layout} shows the layout of the SAE stations currently deployed.\par
This contribution presents the data processing pipeline developed for the SAE scintillation detectors in section~\ref{calib}. It also outlines the latest calibration developments following previous work~\cite{shefali2024statusplansinstrumentationicecube} (Sec.~\ref{calib}), along with an initial assessment of the reconstruction performance benchmarked with respect to IceTop data (Sec.~\ref{performance}) for a selected dataset (Sec.~\ref{data}). In particular, results from Station~0 using data from January-November 2023 are discussed. Finally, in section~\ref{signalmodel} we motivate the use of a data-driven method to model the signal fluctuations in scintillation detectors for potential improvements in the air-shower reconstruction with scintillation detectors. Details involving the status and analysis involving radio antennas of SAE can be found in~\cite{MeghaProceeding, ParasProceeding}.

\label{intro}

\section{Scintillator Data Processing and Calibration}
Each scintillation detector of the SAE is instrumented with a silicon photomultiplier (SiPM), which serves as a photosensor to collect the charge deposited by minimally ionizing particles~(MIPs) traversing the detector. The SiPM signal is digitized using a custom microprocessor-based readout board called uDAQ, which provides three gain channels to enable a wide dynamic range. Both the uDAQ firmware and the scintillation detector data processing algorithms have undergone continuous development since the initial prototype deployment that was carried out in January 2020~\cite{oehler2021development}.

\begin{wrapfigure}{l}{0.45\textwidth}
  \centering
  \includegraphics[width=0.9\linewidth]{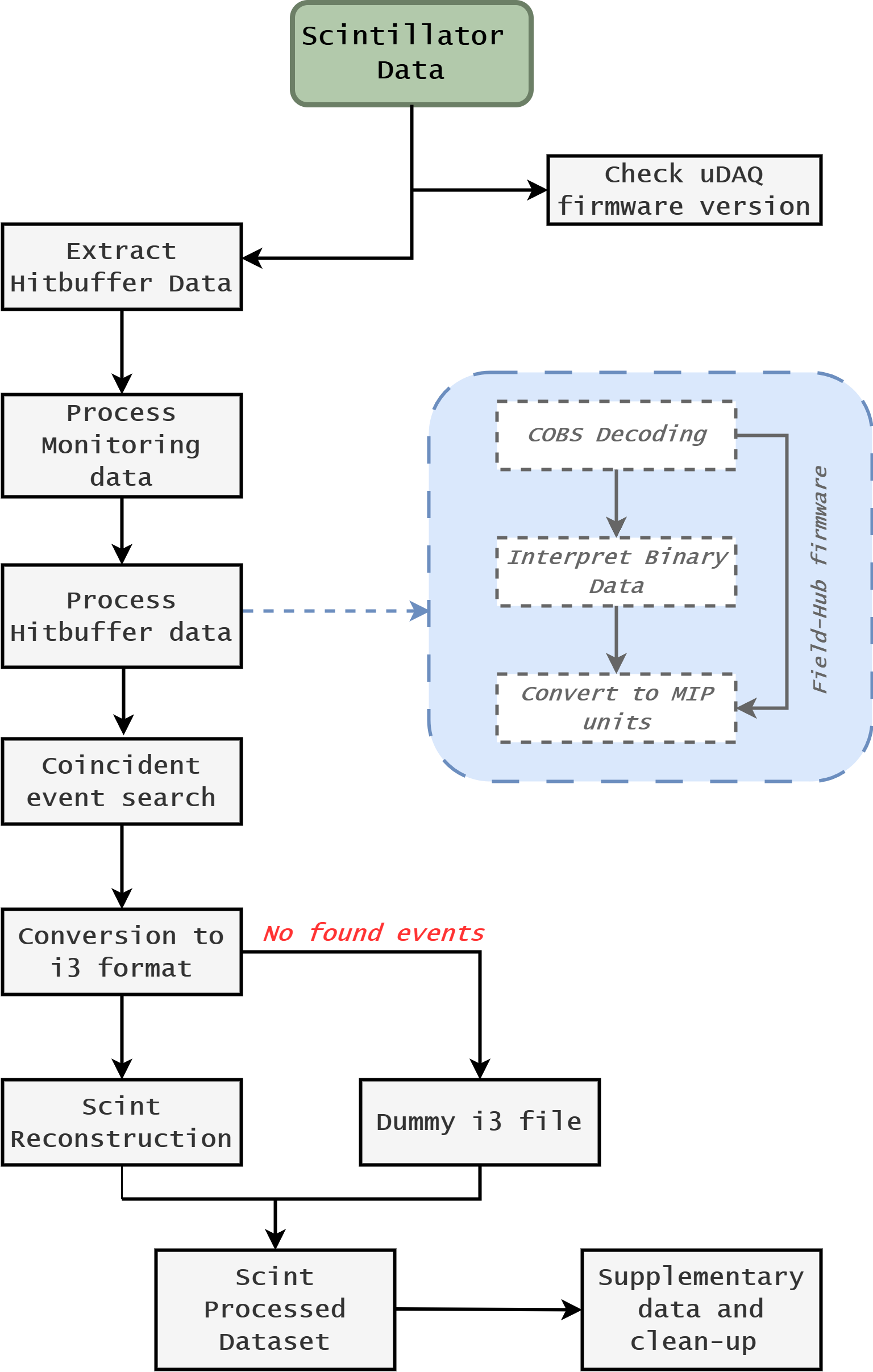}
  \caption{Flowchart describing the scintillator data processing. The data obtained from the air-shower runs is calibrated using the monitoring information and calibration run information. The pipeline also includes reconstruction of the air-showers observed with scintillation detectors after meeting a defined coincidence condition.}
  \label{scint_flow}
\end{wrapfigure}

To ensure compatibility with evolving hardware and firmware, a scalable data processing framework has been implemented to handle multi-year data from the SAE prototype station~\cite{Shefali2025}. The workflow of the processing framework is illustrated in Fig.~\ref{scint_flow}. Within the scope of this work, the scintillator processing module has been further developed to support multiple firmware generations, including a major update introduced in August 2022. This firmware update replaced the previous sequential readout approach with a more efficient system, allowing a significantly higher up-time (up to 23 hours/day) through enhanced buffer capacity and increased readout throughput. The 1 hour is utilized for transferring the data from central DAQ to the ICL machine. However, this improvement was accompanied by a reduction in temperature monitoring resolution, with only a single temperature readout available per run and a significant change in the data format.\par 
The framework facilitates the transformation of raw waveform data from the scintillation detectors (hitbuffer data extraction step in the flowchart) into calibrated physical units, expressed in units of MIPs corresponding to the charge deposited by a single-passing vertical MIP. The calibration stage, highlighted in blue in Fig.~\ref{scint_flow}, includes temperature-dependent gain correction in the high-gain channel, as the SiPM gain exhibits sensitivity to ambient temperature variations (illustrated in~\cite{shefali2024statusplansinstrumentationicecube}).  Furthermore, the MIP peak calibration in the medium- and low-gain channels is performed to obtain the scaling factors necessary for signal saturation correction. These scaling factors enable the recovery of accurate charge measurements by switching to the medium-gain channel when the high-gain channel saturates, and subsequently to the low-gain channel if saturation occurs in the medium-gain channel. Motivated by calibration of IceTop data~\cite{icetop_paper}, for each gain channel, the MIP distribution is modeled using a convolution of a Landau distribution (describing energy loss fluctuations) with an exponential function (to account for the dark count~\footnote{Dark counts correspond to the pulses generated in the SiPM in the absence of light, indicating thermal noise} and background).

\begin{figure}[h]
  \centering
  \includegraphics[width=\textwidth]{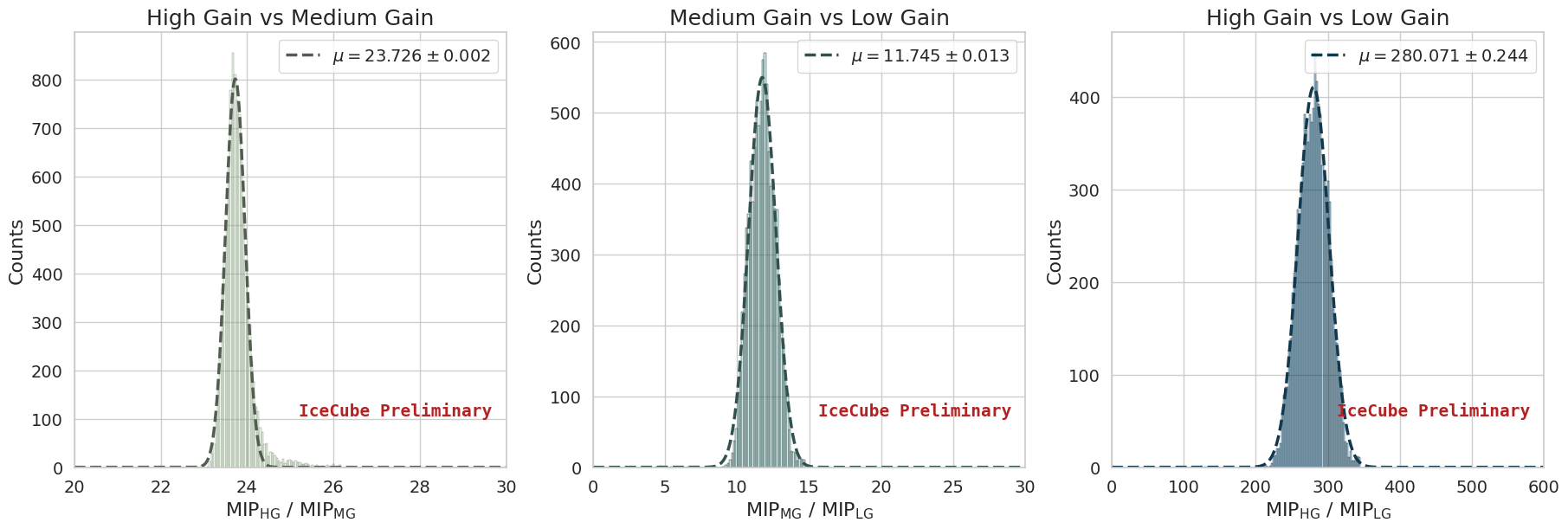}
  \caption{A histogram of the scaling factors for the three gain channels implemented in the uDAQs obtained from calibration data.}
  \label{scaling_factors}
  \vspace*{-0.3cm}
\end{figure}

During the commissioning phase, dedicated calibration measurements were conducted to determine the temperature dependence and scaling between different gain channels. Using data collected between January and June 2023, the gain scaling factors corresponding to uDAQ version 4.1a were derived by obtaining the MIP peak positions from the Landau fits in the three gain channels. A histogram of the scaling factors obtained from these measurements is presented in Fig.~\ref{scaling_factors}. A Gaussian fit is applied to obtain the scaling factors used for the calibration. The low- to high-gain scaling factor is $\sim 280$, which is an indicator of the capability of the detectors' dynamic range.

\begin{wrapfigure}{r}{0.55\textwidth}
  \vspace{-1cm}
  \centering
  \includegraphics[scale=0.37]{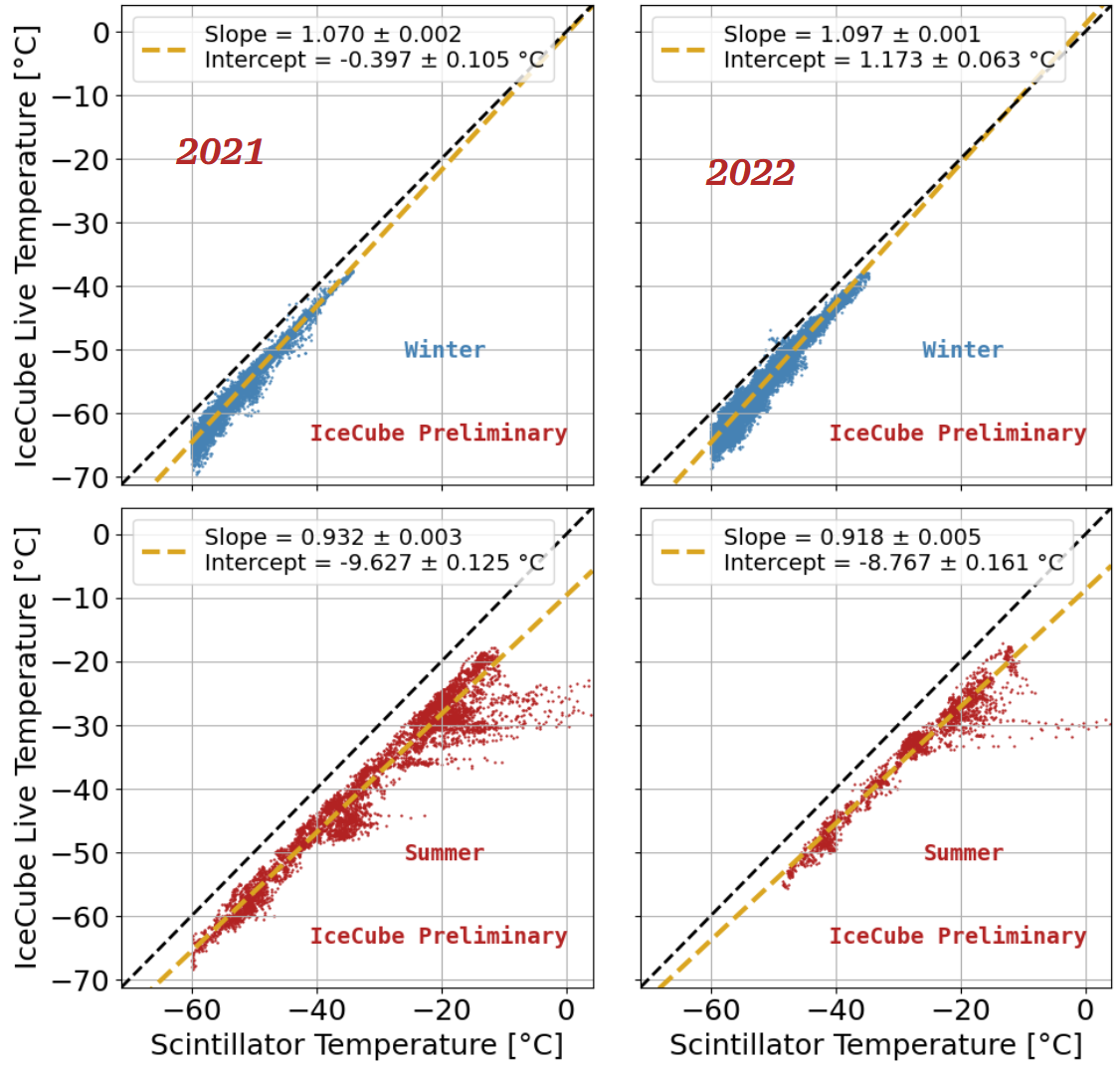}
  \caption{Temperature Correlation Between Scintillation Detectors and ARO Station (2021–Mid 2022, Summer vs. Winter). The data presented is from 2021 to mid-2022.}
  \label{temp_corr}
  \vspace{-0.7cm}
\end{wrapfigure}

As mentioned already, a critical element of scintillator data calibration is temperature calibration, as the response of scintillators is temperature-dependent. Due to the limited temperature information following the firmware upgrade in 2022, a correlation study was performed using temperature data from the South Pole Atmospheric Research Observatory (ARO) weather station~\cite{NOAA_SPO}, which is utilized for the IceCube Live monitoring. Comparing scintillation detectors' internal temperature sensor data from 2021–2022 (obtained with older firmware) with IceCube Live data revealed a strong correlation year round, with an offset of approximately $\sim9^\circ$ in the summer period. Fig.~\ref{temp_corr} presents the correlation observed between the scintillation detectors' data and the IceCube Live data, for the summer~(Oct-Mar) and winter~(Apr-Sept) months of 2021 and 2022, respectively. The deviations observed in summer are likely due to the sun's position relative to the location of Station 0. To reduce the impact of outliers, a Random Sample Consensus (RANSAC) algorithm was applied, providing a robust linear fit for both the slope and intercept. This seasonal offset has been incorporated into the temperature calibration procedure, ensuring consistent performance across different environmental conditions. As a result, the calibration framework supports both single- and multi-station deployments, regardless of firmware version, and remains robust against temperature-induced variations.


\label{calib}

\section{Air Shower dataset}
\vspace{-0.2cm}
After processing, the reconstructed air showers from the scintillator data are matched with IceTop events based on temporal coincidences. For the dataset used, a total of 279,280 events were identified, each observed with six or more scintillators. To evaluate the performance of the prototype station relative to the benchmark detector, IceTop, a subset of these events was selected using the following criteria:
\begin{itemize}[label=--, itemsep=0.5pt, topsep=1pt]
\item Both the scintillator and IceTop reconstruction methods were required to successfully reconstruct the event.
\item Events with a scintillator multiplicity of 8 were selected. This choice is due to the compact size of the prototype station with limited detectors compared to the full IceTop array. Moreover, the reconstruction method involves timing and lateral distribution minimization over up to five parameters, which is best constrained with this multiplicity.
\item Only events with reconstructed IceTop zenith angle up to 37$^\circ$ were retained, where reconstruction is more reliable.
\item A containment cut was applied requiring the IceTop-reconstructed core to lie within $250\mathrm{m}$ of the central DAQ (fieldhub) of the prototype station. This minimizes the number of uncontained showers while maintaining a large enough area to include higher energy showers.
\item Events with an IceTop-reconstructed $log_{10}(S_{125}/\mathrm{VEM})$ \footnote{This is the signal strength at 125\,m distance from the core, and acts as an energy proxy for the primary particle.} greater than 0.5 were retained to ensure maximum reconstruction efficiency and resolution from IceTop.
\item A final zenith angle cut was applied on the scintillator-reconstructed showers, selecting only those with zenith angles below 70$^\circ$, in order to reject misreconstructed events.
\end{itemize}

\begin{wrapfigure}{r}{0.5\textwidth}
\vspace*{-1.7cm}
\centering
\includegraphics[scale=0.38]{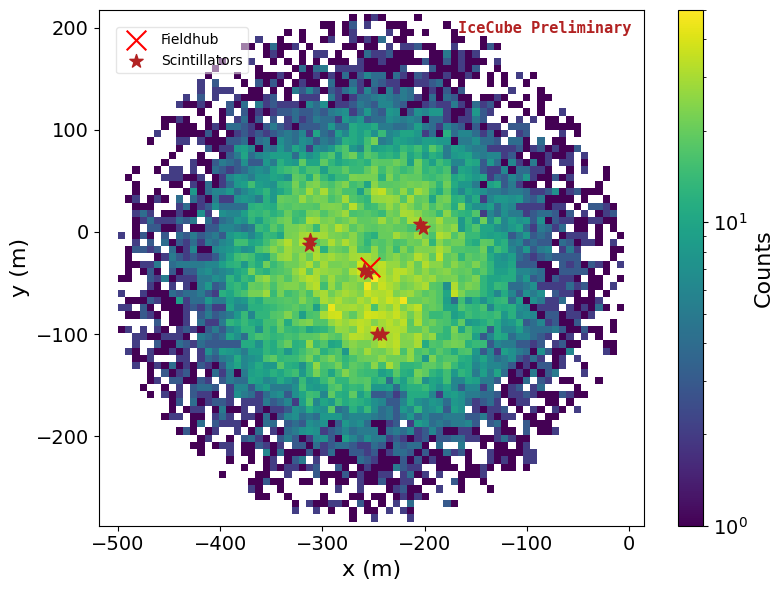}
\vspace*{-0.9cm}
\caption{Distribution of reconstructed shower cores overlaid with the scintillator and fieldhub positions.}
\label{cores}
\vspace*{-0.6cm}
\end{wrapfigure}

Applying these criteria resulted in a final sample of 27,582 events. The distribution of the reconstructed core positions of these events using IceTop reconstruction, along with the locations of the scintillators and the fieldhub, is shown in Fig.~\ref{cores}.

\label{data}

\section{Station 0: Performance}
\begin{figure}[h]
  \centering
  \includegraphics[scale=0.35]{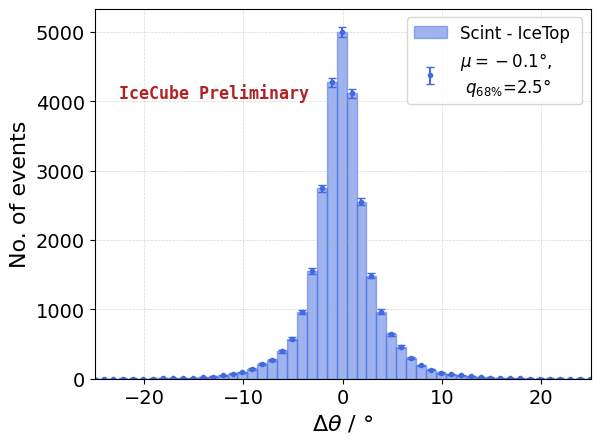}
  \includegraphics[scale=0.35]{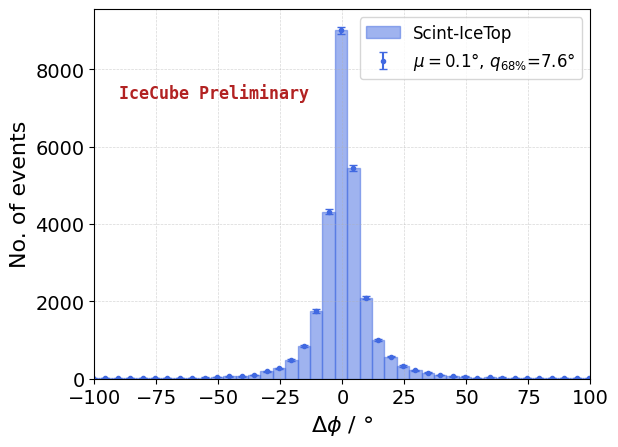}
  \includegraphics[scale=0.35]{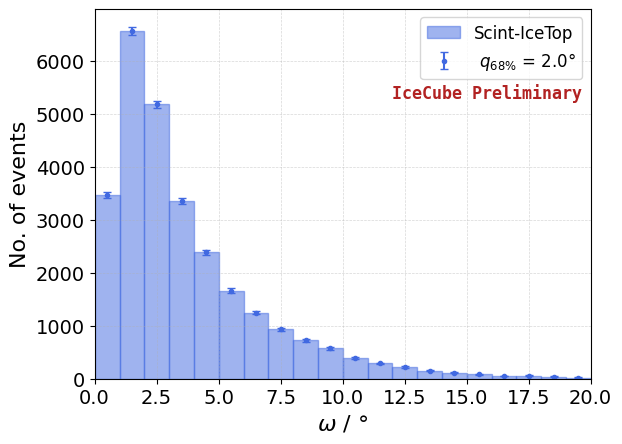}
  \caption{Angular resolution comparison between scintillator and IceTop reconstructions. Top Left, Top Right, and Bottom figures show the distribution of differences in zenith angle, azimuth angle, and total angular distance, respectively.}
  \label{ang_res}
  \vspace*{-0.3cm}
\end{figure}

For the selected events, the performance of the scintillator reconstruction is deduced using the angular resolutions. To assess the angular resolution of the scintillator reconstruction, a comparison is made with IceTop, which serves as the benchmark detector due to its well-established reconstruction performance. The 68th percentile of the distributions is used as a measure of resolution. This is done separately for the zenith angle, azimuth angle, and solid angle, which is the total angular distance between the two reconstructions, and is deduced between the reconstructed directions from the scintillator and IceTop to get the final angular resolution. 
The respective angular resolutions are presented in Fig.~\ref{ang_res}. The combined zenith angular resolution between the IceTop and scintillator reconstruction is observed to be $2.5^\circ$, while for the azimuthal reconstruction it is $7.6^\circ$. This yields a solid angular resolution of $2^\circ$. For near-vertical showers, small differences in direction can lead to large changes in azimuth. This happens because near the vertical ($\theta\approx0^\circ$), the azimuth angle becomes very sensitive and less meaningful due to the geometry of spherical coordinates. Therefore, the solid angular resolution can be used as a good measure of the performance metric.

\begin{figure}[!htb]
  \centering
    \vspace{-0.5cm}
    \includegraphics[scale=0.44]{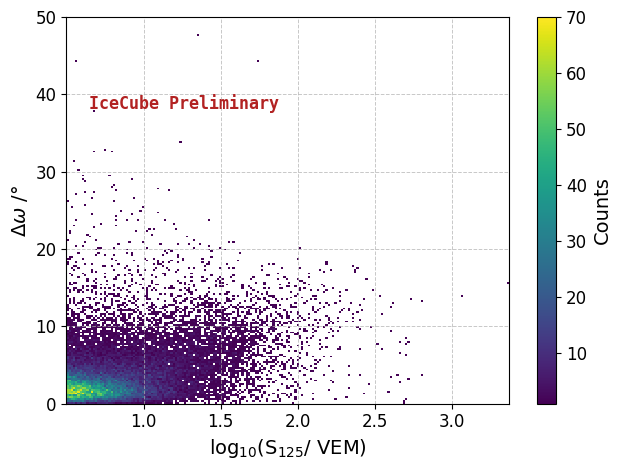}
    \includegraphics[scale=0.37]{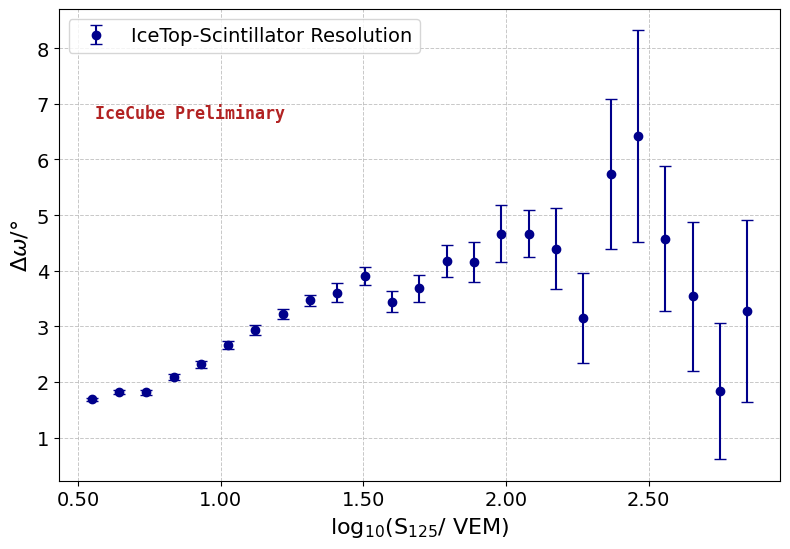}
  \caption{Angular Resolution as a function of $S_{125}$. Left: 2D distribution of the solid angle between scintillator and IceTop reconstructions as a function of $S_{125}$. Right: the 68th percentile of the solid angle in bins of $S_{125}$.}
  \label{omega_vs_energy}
  \vspace*{-0.3cm}
\end{figure}

The solid angular resolution is further examined as a function of the shower size, represented by the IceTop-reconstructed signal at $125\mathrm{m}$ from the core, $S_{125}$. The results are shown in Fig.~\ref{omega_vs_energy}. Figure.~\ref{omega_vs_energy} (left) presents a 2D histogram of the angular resolution ($\Delta\omega$) between the scintillator and IceTop reconstructions as a function of  $log_{10}(S_{125}/VEM)$. Due to the compact footprint of the prototype station, it predominantly detects lower-energy showers, which is reflected in the concentration of events at low $S_{125}$ values. Figure~\ref{omega_vs_energy} (right) shows the binned angular resolution as a function of $S_{125}$. The resolution is estimated using the 68th percentile of the $\Delta \omega$ distribution in each energy bin, with uncertainties derived from bootstrapped resampling. The angular resolution is observed to be roughly $1.8^\circ$ for $log_{10}(S_{125}/\mathrm{VEM}) \lesssim 0.7$, and then gradually increases, plateauing around $4^\circ$. This behavior is consistent with expectation, as higher-energy showers have larger footprints that extend beyond the sensitive area of the single station, reducing reconstruction accuracy. In addition, the dataset also includes lower-energy showers with cores lying outside the station footprint. The presented resolution includes the intrinsic uncertainty of IceTop, and is therefore not yet optimal. It can be further improved through data-driven parametrization of the timing, signal fluctuations and lateral distribution functions specific to the prototype station.

\label{performance}

\section{Signal Model}
\begin{wrapfigure}{r}{0.45\textwidth}
  \centering
  \vspace{-0.7cm}
  \includegraphics[scale=0.32]{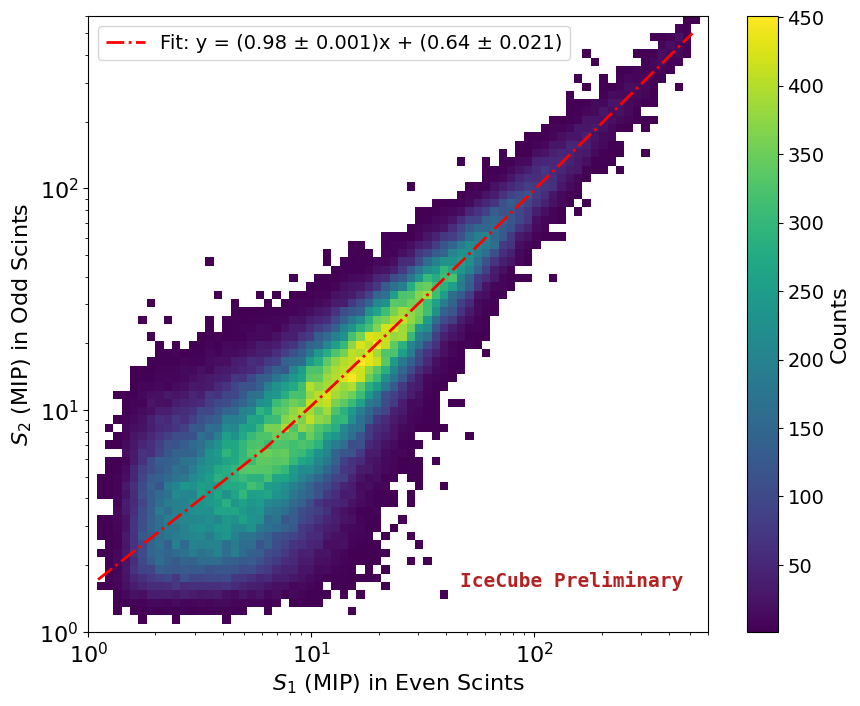}
    \vspace{-0.7cm}
  \caption{Signal deposited in neighboring scintillators. }
  \label{charge_neighbours}
  \vspace*{-0.5cm}
\end{wrapfigure}
\begin{figure}[!htb]
  \centering
  \vspace{-0.2cm}
  \includegraphics[scale=0.39]{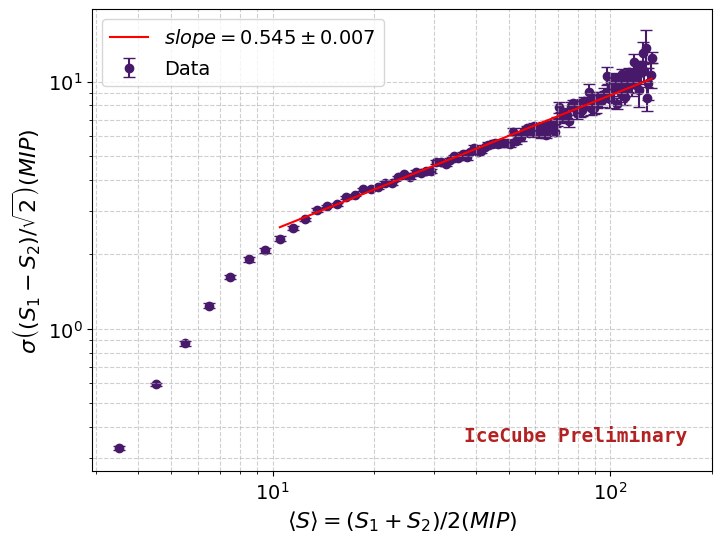}
  \includegraphics[scale=0.39]{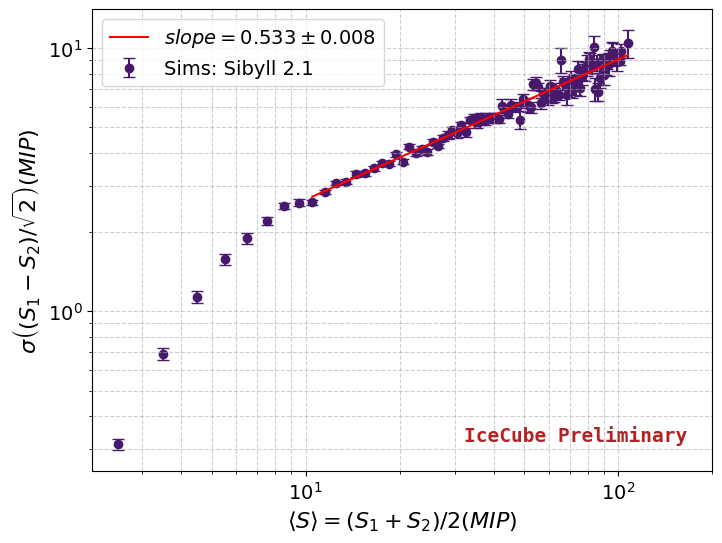}
  \caption{Signal fluctuation as a function of signal size. Left: extracted from data. Right: extracted from single-station simulations. A Poisson-like dependence is observed in both cases.}
  \label{signal_model}
  \vspace*{-0.5cm}
\end{figure}

The performance of the scintillator reconstruction is promising, but further improvements are possible using data-driven parameterization of the detector response as the current simulation-based approach is limited by a simplified detector model. This is explored by studying the signal fluctuations in individual scintillation detectors using the same dataset described above. While the reconstruction previously modeled these fluctuations based on full-array simulations of the SAE, they are now extracted directly from data by comparing signals in neighboring detector pairs of Station~0. To improve statistics, the data from the 4 neighboring detectors placed at 5\,m  was combined into a single distribution, shown in Fig.~\ref{charge_neighbours}. An additional cut was applied to remove detectors within 50\,m of the reconstructed shower core, to avoid the steep gradient of the LDF close to the core, which introduces additional fluctuations which are not due to the detectors themselves. The tight clustering around the diagonal indicates a good relative calibration between detectors during the processing stage.
To extract the fluctuation behavior as a function of signal size, diagonal binning was applied, and the spread (standard deviation of the Gaussian fits) within each bin was computed. The result is shown in Fig.~\ref{signal_model} (left). A rapid fall-off of the fluctuations is visible for mean signals below about 9\,MIP, which is expected due to threshold effects~\cite{AVE2007180}. This behavior is also seen in the charge distribution. Therefore, for the fit, only bins with mean signals above 9\,MIP were used. A power-law fit gives a slope of 0.545, indicating that the fluctuations scale approximately with the square root of the signal, indicating that they follow a Poisson-like behavior. This is in line with theoretical expectations, since the energy deposit in a scintillator arises from a discrete number of particles undergoing energy loss and scattering, which leads to Poissonian fluctuations~\cite{AVE2007180}. 

This data-driven behavior was also reproduced in dedicated single-station simulations of the prototype geometry, shown in Fig.~\ref{signal_model} (right), which rendered a slope of 0.533. Further study of the zenith dependence of these fluctuations is ongoing and will be incorporated into future iterations of the reconstruction.

\label{signalmodel}

\section{Summary and Outlook}
\vspace{-0.2cm}
In this contribution, the calibration and air-shower data preparation of the scintillation detectors of the SAE were presented. Furthermore, the angular resolution relative to the IceTop array was studied, showing a performance of better than 2$^\circ$ for low-energy showers, which is quite promising for a station of such a compact size. A signal fluctuation model, to improve scintillator reconstruction, exhibiting a Poisson dependence, was motivated by data and validated by simulations, and its implementation in the reconstruction framework is planned for the near future.

\bibliographystyle{ICRC}
\bibliography{ICRC2025templates/ICRC2025_template_IceCube}

\clearpage

\section*{Full Author List: IceCube Collaboration}

\scriptsize
\noindent
R. Abbasi$^{16}$,
M. Ackermann$^{63}$,
J. Adams$^{17}$,
S. K. Agarwalla$^{39,\: {\rm a}}$,
J. A. Aguilar$^{10}$,
M. Ahlers$^{21}$,
J.M. Alameddine$^{22}$,
S. Ali$^{35}$,
N. M. Amin$^{43}$,
K. Andeen$^{41}$
% ...
\\
\\
% ...
$^{10}$ Universit{\'e} Libre de Bruxelles, Science Faculty CP230, B-1050 Brussels, Belgium \\
% ...
$^{16}$ Department of Physics, Loyola University Chicago, Chicago, IL 60660, USA \\
$^{17}$ Dept. of Physics and Astronomy, University of Canterbury, Private Bag 4800, Christchurch, New Zealand \\
% ...
$^{21}$ Niels Bohr Institute, University of Copenhagen, DK-2100 Copenhagen, Denmark \\
$^{22}$ Dept. of Physics, TU Dortmund University, D-44221 Dortmund, Germany \\
% ...
$^{35}$ Dept. of Physics and Astronomy, University of Kansas, Lawrence, KS 66045, USA \\
% ...
$^{39}$ Dept. of Physics and Wisconsin IceCube Particle Astrophysics Center, University of Wisconsin{\textemdash}Madison, Madison, WI 53706, USA \\
% ...
$^{41}$ Department of Physics, Marquette University, Milwaukee, WI 53201, USA \\
% ...
$^{43}$ Bartol Research Institute and Dept. of Physics and Astronomy, University of Delaware, Newark, DE 19716, USA \\
% ...
$^{63}$ Deutsches Elektronen-Synchrotron DESY, Platanenallee 6, D-15738 Zeuthen, Germany \\
$^{\rm a}$ also at Institute of Physics, Sachivalaya Marg, Sainik School Post, Bhubaneswar 751005, India \\
% ...

\subsection*{Acknowledgments}

\noindent
The authors gratefully acknowledge the support from the following agencies and institutions:
USA {\textendash} U.S. National Science Foundation-Office of Polar Programs,
U.S. National Science Foundation-Physics Division,
U.S. National Science Foundation-EPSCoR,
U.S. National Science Foundation-Office of Advanced Cyberinfrastructure,
Wisconsin Alumni Research Foundation,
Center for High Throughput Computing (CHTC) at the University of Wisconsin{\textendash}Madison,
Open Science Grid (OSG),
Partnership to Advance Throughput Computing (PATh),
Advanced Cyberinfrastructure Coordination Ecosystem: Services {\&} Support (ACCESS),
Frontera and Ranch computing project at the Texas Advanced Computing Center,
U.S. Department of Energy-National Energy Research Scientific Computing Center,
Particle astrophysics research computing center at the University of Maryland,
Institute for Cyber-Enabled Research at Michigan State University,
Astroparticle physics computational facility at Marquette University,
NVIDIA Corporation,
and Google Cloud Platform;
Belgium {\textendash} Funds for Scientific Research (FRS-FNRS and FWO),
FWO Odysseus and Big Science programmes,
and Belgian Federal Science Policy Office (Belspo);
Germany {\textendash} Bundesministerium f{\"u}r Forschung, Technologie und Raumfahrt (BMFTR),
Deutsche Forschungsgemeinschaft (DFG),
Helmholtz Alliance for Astroparticle Physics (HAP),
Initiative and Networking Fund of the Helmholtz Association,
Deutsches Elektronen Synchrotron (DESY),
and High Performance Computing cluster of the RWTH Aachen;
Sweden {\textendash} Swedish Research Council,
Swedish Polar Research Secretariat,
Swedish National Infrastructure for Computing (SNIC),
and Knut and Alice Wallenberg Foundation;
European Union {\textendash} EGI Advanced Computing for research;
Australia {\textendash} Australian Research Council;
Canada {\textendash} Natural Sciences and Engineering Research Council of Canada,
Calcul Qu{\'e}bec, Compute Ontario, Canada Foundation for Innovation, WestGrid, and Digital Research Alliance of Canada;
Denmark {\textendash} Villum Fonden, Carlsberg Foundation, and European Commission;
New Zealand {\textendash} Marsden Fund;
Japan {\textendash} Japan Society for Promotion of Science (JSPS)
and Institute for Global Prominent Research (IGPR) of Chiba University;
Korea {\textendash} National Research Foundation of Korea (NRF);
Switzerland {\textendash} Swiss National Science Foundation (SNSF).

\end{document}